\documentclass[prb,aps,twocolumn,amsmath,amssymb,showpacs]{revtex4}
\usepackage{graphicx}	
\DeclareGraphicsExtensions{.png,.eps}
\usepackage[english]{babel}
\usepackage{natbib}
\usepackage{amsfonts}
\usepackage{amssymb}
\usepackage{dcolumn}
\usepackage{amsmath}
\usepackage[ansinew]{inputenc}
\usepackage{amssymb}
\usepackage{multirow}
\usepackage{mathtools}\begin{tiny}\end{tiny}
\usepackage{wasysym}
\usepackage{color}
\usepackage{verbatim}
\usepackage{ulem}

\renewcommand{\vec}[1]{{\bf #1}}

\newcommand{\old}[1]{}

\begin{document}
\date{\today}

\title{Deviation From the Landau-Lifshitz-Gilbert equation in the Inertial regime of the Magnetization}

\author{E. Olive}
\author{Y. Lansac}
\affiliation{GREMAN, UMR 7347, Universit\'e Fran\c cois Rabelais-CNRS, Parc de Grandmont, 37200 Tours, France}
\author{M. Meyer}
\author{M. Hayoun}
\author{J.-E. Wegrowe}
\affiliation{Laboratoire des Solides Irradi\'es, \'Ecole Polytechnique, CEA-DSM,  CNRS, F-91128 Palaiseau, France}

\begin{abstract}
We investigate in details the inertial dynamics of a uniform magnetization in the ferromagnetic resonance (FMR) context. Analytical predictions and numerical simulations of the complete equations within the Inertial Landau-Lifshitz-Gilbert (ILLG) model are presented. In addition to the usual precession resonance, the inertial model gives a second resonance peak associated to the nutation dynamics provided that the damping is not too large. 
The analytical resolution of the equations of motion yields both the precession and nutation angular frequencies. They are function of the inertial dynamics characteristic time $\tau$, the dimensionless damping $\alpha$ and the static magnetic field $H$. A scaling function with respect to $\alpha\tau\gamma H$ is found for the nutation angular frequency, also valid for the precession angular frequency when $\alpha\tau\gamma H\gg 1$. Beyond the direct measurement of the nutation resonance peak, we show that the inertial dynamics of the magnetization has measurable effects on both the width and the angular frequency of the precession resonance peak when varying the applied static field. These predictions could be used to experimentally identify the inertial dynamics of the magnetization proposed in the ILLG model.
\end{abstract}

\maketitle

\section{Introduction}
\label{Intro}
The Landau-Lifshitz-Gilbert (LLG) equation is a kinetic equation that does not contain acceleration terms, i.e. that does not contain inertia. The corresponding trajectory is reduced to a damped precession around the axis defined by the effective field. The measurement of this precession is usually performed by the mean of ferromagnetic resonance (FMR). The power absorbed by the system is then measured at steady state while adding an oscillatory field to the effective field, and tuning the frequency close to the resonance frequency.
However, the validity of the LLG equation is limited to large time scales \cite{Brown}, or low frequency regimes (similarly to the Debye model of electric dipoles \cite{Kubo}). Indeed, the precession with damping described by the LLG equation is a diffusion process in a field of force, for which the angular momentum has reached  equilibrium. Accordingly, if the measurements are performed at fast enough time scales, or high enough frequencies, inertial terms should be expected to play a role in the dynamics, which is no longer reduced to a damped precession \cite{PhD,inertia,Fahnle,AJP,APL,Bott,Bhat}. A nutation dynamics is therefore expected, giving a second resonant peak at the nutation frequency, and this new absorption should be measurable with dedicated spectroscopy (e.g. using infrared spectroscopy). \\
Despite its fundamental importance, a systematic experimental investigation of possible inertial effects of the uniform magnetization has however been overlooked.
In order to evidence experimentally the consequences of inertia in the dynamics of a uniform magnetization, it is first necessary to establish the characteristics that would allow to discriminate inertia from spurious effects in spectroscopy experiments.
We propose in this paper some simple theoretical and numerical tools than can be used by experimentalists in order to evidence unambiguously the effects of inertia of the magnetization.\\

 The LLG equation reads :
\begin{equation}
\frac{d \vec{\bf M}}{dt} = \gamma \vec{\bf M} \times \left[\vec{\bf H}^\textrm{eff} - \eta \frac{d \vec{\bf M}}{dt} \right]
\label{eq1}
\end{equation}
where $\bf M$ is the magnetization, $\bf {H}^\textrm{eff}$ the effective magnetic field, $\eta$ the Gilbert damping, and $\gamma$ the gyromagnetic ratio.
If the description is extended to the fast degrees of freedom (i.e. the degrees of freedom that includes the time derivative of the angular momentum), a supplementary inertial term should be added with the corresponding relaxation time $\tau$. From this Inertial Landau-Lifshitz-Gilbert (ILLG) model, the new equation reads \cite{PhD,inertia,Fahnle,AJP,APL} :
\begin{equation}
\frac{d \vec M}{dt} = \gamma \vec M \times \left[\vec H^\textrm{eff} - \eta \left(\frac{d \vec M}{dt} + \tau \frac{d^2 \vec M}{dt^2} \right) \right] 
\label{eq2}
\end{equation}

One of the main consequences of the new dynamical equation is the emergence of the second resonance peak associated to the nutation at high frequencies, as reported in our previous study \cite{APL}. In the literature the nutation dynamics of magnetic moments has been investigated using various theoretical approaches though not yet evidenced experimentally. B\"ottcher and Henk studied the significance of nutation in magnetization dynamics of nanostructures such as a chain of Fe atoms, and Co islands on Cu(111) \cite{Bott}. 
They found that the nutation is significant on the femtosecond time scale with a typical damping constant of 0.01 up to 0.1. Moreover, they concluded that nutation shows up preferably in low-dimensional systems but with a small amplitude with respect to the precession.
Zhu {\it et al.} predicted a nutation dynamics for a single spin embedded in the tunneling barrier between two superconductors \cite{Zhu}. This unusual spin dynamics is caused by coupling to a Josephson current. They argue that this prediction might be directly tested for macroscopic spin clusters.
The nutation is also involved in the dynamics of a single spin embedded in the tunnel junction between ferromagnets in the presence of an alternating current \cite{Fran}.
In an atomistic framework, Bhattacharjee {\it et al.} showed that first-principle techniques used to calculate the Gilbert damping factor may be extended to calculate the moment of inertia tensor associated to the nutation \cite{Bhat}.

Our previous work \cite{APL} was focussed on the short time nutation dynamics generated by the ILLG equation, and was limited to fixed values of the inertial characteristic time scale $\tau$, the dimensionless damping $\alpha$ and the static field $H$.
In this paper we present a combined analytical and numerical simulation study of the ILLG equation with new results. In particular we derive analytical results in the small inclination limit that can be used in ferromagnetic resonance (FMR) experiments, and which allow to predict both the precession and nutation resonance angular frequencies. We also investigate the ILLG equation while varying the three parameters $\alpha$, $\tau$ and $H$, and scaling functions are found. Finally, we present important indications for experimental investigations of the inertial dynamics of the magnetization. Indeed, a consequence of the ILLG equation is the displacement of the well-known FMR peak combined with a modified shape with respect to that given by the LLG equation. This displacement could not be without consequences on the determination of the gyromagnetic factor $\gamma$ by ferromagnetic resonance.

The paper is organized as follows. In section \ref{Analytique} we show analytical solutions of the precession and nutation dynamics for the uniform magnetization in a static applied field $\bf H$. The small inclination limit is investigated in order to reproduce the usual experimental FMR context.
In section \ref{Simul} we describe the numerical simulations of the magnetization inertial dynamics in both a static and a small perpendicular sinusoidal magnetic field ($\bf H^\textrm{eff} = \bf H + \bf h_{\perp}(\omega)$). The resonance curves are computed and, provided that the damping is not too large, a nutation resonance peak appears in addition to the usual ferromagnetic resonance peak associated to the magnetization precession. In section \ref{Results} the behavior of the ILLG equation is investigated in details while varying the characteristic time $\tau$ of the inertial dynamics, the dimensionless damping $\alpha$ and the static field $H$. A very good agreement is found between the analytical and numerical simulation results, and a scaling function with respect to $\alpha\tau\gamma H$ is found. In section \ref{Experiments} we propose experiments in the FMR context that should evidence the inertial dynamics of the magnetization described in the ILLG model. In particular, when the static field is varied, the ILLG precession resonance peak has different behaviors compared to the usual LLG precession peak with shifted resonance angular frequency and modified shape. We show that the differences between LLG and ILLG precession peaks are more pronounced in large damping materials and increase with the static field.
Finally, we derive the conclusions in section \ref{Conclusion}.

\section{Analytical solutions for the ILLG equation}
\label{Analytique}

The magnetization position is described in spherical coordinates $(M_\textrm{s},\theta,\phi)$, where $M_\textrm{s}$ is the  radius coordinate fixed at a constant value for the uniformly magnetized body, $\theta$ is the inclination and $\phi$ is the azimuthal angle. In a static magnetic field $H \vec{\hat{z}}$ applied in the $z$ direction,  {\it i.e.} $\vec H=H \left(\cos \theta\ \vec e_r - \sin \theta\ \vec e_{\theta}\right)$ in the spherical basis $\left(\vec e_r,\vec e_{\theta},\vec e_{\phi} \right)$, Eq. (\ref{eq2}) gives the following system :
\begin{subequations}\label{eq3}
\begin{eqnarray}
\nonumber
\ddot\theta & = & -\frac{1}{\tau} \dot\theta - \frac{1}{\tau_1}\dot\phi \sin\theta + \dot\phi^2 \sin\theta \cos\theta  \\
 & & - \frac{\omega_2}{\tau_1} \sin\theta \\
\ddot\phi \sin\theta & = & \frac{1}{\tau_1}\dot\theta - \frac{1}{\tau} \dot\phi \sin\theta - 2\dot\phi \dot\theta \cos\theta 
\end{eqnarray}
\end{subequations}
where the characteristic times are $\tau$ and $\tau_1= \alpha \tau$, 
$\omega_2 = \gamma H$ is the Larmor angular frequency, 
and $\alpha=\gamma \eta M_\textrm{s}$ is the dimensionless damping.

Using the dimensionless time $t'=t/\tau$, Eqs. (\ref{eq3}) become
\begin{subequations}\label{eq4}
\begin{eqnarray}
\nonumber
\theta'' & = & - \theta' - {\widetilde\tau_1}\phi' \sin\theta + \phi'^2 \sin\theta \cos\theta \\
& & - {\widetilde\omega_2\widetilde\tau_1} \sin\theta 
 \\
\nonumber
& & \\
\phi'' \sin\theta & = & {\widetilde\tau_1}\theta' - \phi' \sin\theta - 2\phi' \theta' \cos\theta \\
\nonumber
\end{eqnarray}
\end{subequations}
where 
$$\theta'=d\theta/dt',\ \theta''=d^2\theta/dt'^2,\ \phi'=d\phi/dt',\ \phi''=d^2\phi/dt'^2,$$
and 
\begin{eqnarray*}
{\widetilde\tau_1} & = & \frac{\tau} {\tau_1} = \frac{1} {\alpha}\\    
{\widetilde\omega_2} & = & {\omega_2\tau} = \tau \gamma H \\ 
\end{eqnarray*}

In the following subsections we extract analytical results that can be used to predict the positions in the angular frequency domain of the precession and nutation resonance peaks. We will consider the small inclination limit which holds in the FMR context.

\subsection{Precession : exact and approximate solutions}

To determine the precession dynamics of the inertial model we search for the long time scale solution $\phi'(t')=\phi'_\textrm{prec}$, where $\phi'_\textrm{prec}$ is the constant precession velocity. Since the damping progressively shifts the magnetization to the $z$ axis, we investigate the small inclination limit where $\phi'(t')=\phi'_\textrm{prec}$ should hold. With $\sin\theta\sim\theta$ and $\cos\theta\sim 1$, Eqs. (\ref{eq4}) therefore reads :
\begin{subequations}\label{eq5}
\begin{eqnarray}
\theta'' + \theta'+{\widetilde\omega_0}^2\ \theta =  0\\
\phi'_\textrm{prec} = \frac{{\widetilde\tau_1}\theta' }{\theta + 2\theta'}\\
\nonumber
\end{eqnarray}
\end{subequations}
where the natural angular frequency of the overdamped harmonic oscillator $\theta(t')$ defined by Eq. (\ref{eq5}a) is given by
\begin{equation}\label{eq6}
{\widetilde\omega_0}=\sqrt{{\widetilde\tau_1}(\phi'_\textrm{prec}+{\widetilde\omega_2})-\phi'^{\ 2}_\textrm{prec}}
\end{equation}
The characteristic equation  associated to the differential equation Eq. (\ref{eq5}a) is $\beta^2+\beta+{\widetilde\omega_0}^2=0$ which gives in the aperiodic regime the two solutions \begin{equation}\label{eq7}
\beta_{\pm}=\frac{-1\pm\sqrt{1-4{\widetilde\omega_0}^2}}{2}
\end{equation}
Since $|\beta_+|<|\beta_-|$, the inclination of the magnetization behaves at long time scales as $$\theta(t')\sim e^{\beta_{\textrm{+}}t'}\ ,$$
which inserted in Eq. (\ref{eq5}b) gives
\begin{equation}\label{eq8}
\phi'_\textrm{prec} = \frac{{\widetilde\tau_1} \beta_+}{1 + 2\beta_+}
\end{equation}
In original time units, the precession velocity $\dot\phi_\textrm{prec}$ is therefore the solution of 
\begin{equation}\label{eq9}
\dot\phi_\textrm{prec} = \frac{ \beta_+(\dot\phi_\textrm{prec})}{\alpha\tau \left(1 + 2\beta_+(\dot\phi_\textrm{prec})\right)}
\end{equation}
where the function $\beta_+(\dot\phi_\textrm{prec})$ is given by 
\begin{equation}\label{eq10}
\beta_+(\dot\phi_\textrm{prec})=\frac{-1+\sqrt{1-4\tau\left(\frac{\dot\phi_\textrm{prec}+\gamma H}{\alpha}-\tau\dot\phi^{\ 2}_\textrm{prec}\right)}}{2}
\end{equation}

Equation \ref{eq9} may be numerically solved to extract the precession velocity, and therefore the precession resonance peak when a sinusoidal magnetic field $\bf h_{\perp}(\omega)$ is superimposed perpendicular to the static field $H \vec{\hat{z}}$.\\
 For $\tau\ll 10^{-11}s$ and $\alpha\le 0.1$, the precession velocity $\dot\phi_\textrm{prec}$ for small applied static fields may be accurately evaluated from a quadratic equation : in this case ${\widetilde\omega_0}^2\ll 1$ and Eq. (\ref{eq7}) leads to $\beta_+\approx -{\widetilde\omega_0}^2$. Eq. (\ref{eq8}) therefore gives a cubic equation in $\phi'_\textrm{prec}$ where the cubic term $-2\alpha\phi'^{\ 3}_\textrm{prec}$ is negligeable. In this case the solution of the resulting quadratic equation is in original time units
\begin{equation}\label{eq11}
\dot\phi_\textrm{prec}=\frac{-b-\sqrt{b^2+12\tau\gamma H/\alpha}}{6\tau}
\end{equation}
 with $b=2\tau\gamma H-\alpha-1/\alpha$. We choose the negative solution of the quadratic equation in order to agree with the negative velocity $\dot\phi_\textrm{LLG}=-\gamma H/(1+\alpha^2)$ given by the LLG model.

\subsection{Nutation : angular frequency}

Unlike the precession, the nutation properties should be derived considering intermediate time scales where the precession has not yet reached a constant velocity. Eqs. (\ref{eq4}) should therefore be reconsidered. 
To derive the nutation properties, it is convenient to examine the angular velocity $\theta'$.
For simplicity we note $\theta'={\widetilde\omega_{\theta}}$ and $\phi'={\widetilde\omega_{\phi}}$.  Eqs. (\ref{eq4}) therefore rewrite
\begin{subequations}\label{eq12}
\begin{eqnarray}
\nonumber
{\widetilde\omega_{\theta}}' & = & - {\widetilde\omega_{\theta}} - {\widetilde\tau_1}{\widetilde\omega_{\phi}}\sin\theta + {\widetilde\omega_{\phi}}^2 \sin\theta \cos\theta \\
& & - {\widetilde\omega_2\widetilde\tau_1} \sin\theta 
 \\
\nonumber
& & \\
{\widetilde\omega_{\phi}}' \sin\theta & = & {\widetilde\tau_1}{\widetilde\omega_{\theta}} - {\widetilde\omega_{\phi}} \sin\theta - 2{\widetilde\omega_{\phi}} {\widetilde\omega_{\theta}} \cos\theta \\
\nonumber
\end{eqnarray}
\end{subequations}
We derive Eq. (\ref{eq12}a) with respect to time $t'$ which gives
\begin{eqnarray}
\nonumber
{\widetilde\omega_{\theta}}'' & = & -{\widetilde\omega_{\theta}}'+(2\ {\widetilde\omega_{\phi}} \cos\theta-{\widetilde\tau_1}){\widetilde\omega_{\phi}}'\sin\theta - {\widetilde\tau_1}{\widetilde\omega_{\phi}}{\widetilde\omega_{\theta}}\cos\theta \\
\nonumber
& & +{\widetilde\omega_{\phi}}^2 {\widetilde\omega_{\theta}}(\cos^2\theta-\sin^2\theta)- {\widetilde\omega_2\widetilde\tau_1} {\widetilde\omega_{\theta}}\cos\theta
\end{eqnarray}
where the term ${\widetilde\omega_{\phi}}'\sin\theta$ may be replaced with the expression in Eq. (\ref{eq12}b). We therefore obtain
\begin{eqnarray}\label{eq13}
\nonumber
& & {\widetilde\omega_{\theta}}''  +  {\widetilde\omega_{\theta}}' + \left({\widetilde\tau_1}^2+{\widetilde\omega_2\widetilde\tau_1}\cos\theta\right) {\widetilde\omega_{\theta}} =  \\ \nonumber
& & {\widetilde\tau_1}{\widetilde\omega_{\phi}}\sin\theta + 3\ {\widetilde\tau_1}{\widetilde\omega_{\phi}}{\widetilde\omega_{\theta}}\cos\theta -2\ {\widetilde\omega_{\phi}}^2\cos\theta\sin\theta\\
& &  
- (3\cos^2\theta+\sin^2\theta){\widetilde\omega_{\phi}}^2 {\widetilde\omega_{\theta}}
\end{eqnarray}
Eq. (\ref{eq13}) should be closely related to the nutation dynamics since it describes the ${\widetilde\omega_{\theta}}$ oscillator. This assumption will be confirmed in section \ref{NutationPeakAnalSimul} for a broad range of parameters. Eq. (\ref{eq13})
defines the damped oscillator ${\widetilde\omega_{\theta}}$ which is non-linearly coupled to the ${\widetilde\omega_{\phi}}$ oscillator. This expression shows that, in the absence of coupling and in the small inclination limit $\theta\ll 1\ rad$, the ${\widetilde\omega_{\theta}}$ oscillator oscillates at the natural angular frequency $\sqrt{{\widetilde\tau_1}^2+{\widetilde\omega_2\widetilde\tau_1}}$. We therefore deduce an approximate expression for the nutation angular frequency in the weak coupling case which is given by the expression 
\begin{eqnarray}\label{eq14}
{\widetilde\omega_\textrm{nu}^\textrm{weak}} = \sqrt{{\widetilde\tau_1}^2+{\widetilde\omega_2\widetilde\tau_1}}
\end{eqnarray}
which in original time units gives
\begin{eqnarray}\label{eq15}
{\omega_\textrm{nu}^\textrm{weak}} = \frac{\sqrt{1+\alpha\tau\gamma H}}{\alpha\tau}
\end{eqnarray}
\\
From Eq. (\ref{eq15}) we deduce the following asymptotic behaviors :  when $\tau\ll 1/\alpha\gamma H$ then ${\omega_\textrm{nu}^\textrm{weak}}\sim 1/\alpha\tau$, and when $\tau\gg 1/\alpha\gamma H$ then ${\omega_\textrm{nu}^\textrm{weak}}\sim 1/\sqrt{\alpha\tau}$.\\

Because of the non-linear coupling terms in the right-hand side of Eq. (\ref{eq13}), the true position of the nutation resonance peak in FMR experiments may differ from the approximate angular frequency defined by Eq. (\ref{eq15}).
However the simulation of the resonance curves  with a sinusoidal magnetic field $\bf h_{\perp}(\omega)$  superimposed perpendicular to the static field $H \vec{\hat{z}}$ will show in section \ref{NutationPeakAnalSimul} that the non-linear coupling terms only slightly shift the nutation resonance peak from the approximate angular frequency.

\section{Numerical simulations of the resonance curves in the ILLG model}
\label{Simul}

We apply a fixed magnetic field $\vec H=H \vec{\hat{z}}$ along the $z$ direction, and a small sinusoidal magnetic field $\vec h_{\perp} = h_{\perp} \cos\omega t\ \vec{\hat x}$ in the $x$ direction.  In the spherical basis the components of the total magnetic field $\vec H^{\textrm{eff}} =\vec H + \vec h_{\perp}$ in Eq. (\ref{eq2}) are
\begin{eqnarray*}
H_\textrm{r}^{\textrm{eff}} & = & H  \cos\theta + h_{\perp} \sin\theta \cos\phi \cos\omega t\\
H_{\theta}^{\textrm{eff}} & = & -H  \sin\theta + h_{\perp} \cos\theta \cos\phi \cos\omega t \\
H_{\phi}^{\textrm{eff}} & = & - h_{\perp} \sin\phi \cos\omega t. 
\end{eqnarray*}
which lead to the following dynamical equations for the spherical angles $(\theta,\phi)$ of the magnetization 
\begin{subequations}\label{eq16}
\begin{eqnarray}
\nonumber
\ddot\theta & = & -\frac{1}{\tau} \dot\theta - \frac{1}{\tau_1}\dot\phi \sin\theta + \dot\phi^2 \sin\theta \cos\theta \\  & & - \frac{\omega_2}{\tau_1} \sin\theta  + \frac{\omega_3}{\tau_1} \cos\theta \cos\phi \cos\omega t \\
\nonumber
& & \\
\nonumber
& & \\
\nonumber
\ddot\phi \sin\theta & = & \frac{1}{\tau_1}\dot\theta - \frac{1}{\tau} \dot\phi \sin\theta - 2\dot\phi \dot\theta \cos\theta \\
& & - \frac{\omega_3}{\tau_1} \sin\phi \cos\omega t
\end{eqnarray}
\end{subequations}
where $\omega_3 =\gamma h_{\perp}$ is the angular frequency associated to the sinusoidal field.

 Using the dimensionless time $t'=t/\tau$, Eqs. (\ref{eq16}) become
\begin{subequations}\label{eq17}
\begin{eqnarray}
\nonumber
\theta'' & = & - \theta' - {\widetilde\tau_1}\phi' \sin\theta + \phi'^2 \sin\theta \cos\theta \\
& & - {\widetilde\omega_2\widetilde\tau_1} \sin\theta + {\widetilde\omega_3\widetilde\tau_1}\cos\theta \cos\phi \cos{\widetilde\omega}t' \\
\nonumber
& & \\
\nonumber
\phi'' \sin\theta & = & {\widetilde\tau_1}\theta' - \phi' \sin\theta - 2\phi' \theta' \cos\theta \\
& & - {\widetilde\omega_3\widetilde\tau_1}\sin\phi \cos{\widetilde\omega}t'
\end{eqnarray}
\end{subequations}
where 
$$\theta'=d\theta/dt',\ \theta''=d^2\theta/dt'^2,\ \phi'=d\phi/dt',\ \phi''=d^2\phi/dt'^2,$$
and 
\begin{eqnarray*}
{\widetilde\tau_1} & = & {\tau \over \tau_1} = {1 \over  \alpha}\\    
{\widetilde\omega_2} & = & {\omega_2\tau} = \tau \gamma H \\ 
{\widetilde\omega_3} & = & {\omega_3\tau} = \tau \gamma h_{\perp}\\
{\widetilde\omega} & = & {\omega\tau}  
\end{eqnarray*}

We use $\gamma=10^{11}\ rad . s^{-1}.T^{-1}$, and we vary the characteristic time $\tau$ for three different values of the dimensionless damping $\alpha=0.1$, $0.01$ and $0.5$. We investigate several values of the static magnetic field from $H=0.2\ T$ up to $H=200\ T$.
We numerically integrate Eqs. (\ref{eq17}) using either a double precision second order Runge-Kutta algorithm or a  double precision five  order Gear  algorithm \cite{G1971}. Typically, we use time steps $10^{-7}<\delta t'<10^{-3}$ depending on the values of $\tau$ and $\omega$.

The resonance curves are obtained by investigating the magnetization response to the small oscillating field $\vec h_{\perp}(\omega) = h_{\perp} \cos\omega t\ \vec{\hat x}$ applied perpendicular to the static field $\vec H= H \ \vec{\hat z}$. We analyse the permanent dynamical regime where the magnetization components oscillate around well defined mean values.
For fixed values of the oscillating field angular frequency $\omega$ and oscillating field amplitude $h_{\perp}$, we compute the mean value $<M_{\perp}>$ (averaged over time) of the transverse magnetization  $M_{\perp}(t)=\sqrt {M_x^2(t)+M_y^2(t)}$, from which we extract for fixed values of $\omega$ the transverse susceptibility 
defined by $\chi_{\perp}=d<M_{\perp}>/dh_{\perp}$. 
We choose values of the oscillating field amplitude $h_{\perp}=10^{-1}, 10^{-2}, 10^{-3}, 10^{-4}$ and $10^{-5}\ T$, and we plot $<M_{\perp}>$ with respect to $h_{\perp}$ for each $\omega$. 
As an example, we show the case $\alpha=0.1$, $\tau=2\times 10^{-10}\ s$, $H=2\ T$ and $\omega=1.2\times 10^{11}\ rad.s^{-1}$.
\begin{figure}[!h]
\includegraphics[scale=0.4]{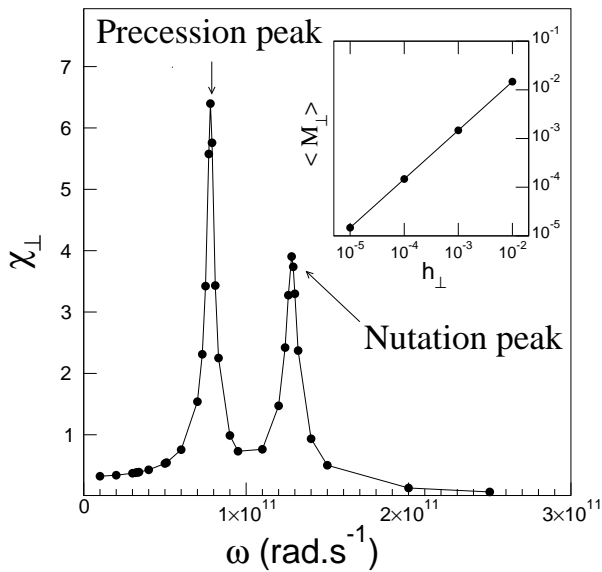}
\caption{Resonance curves of the transverse susceptibility $\chi_{\perp}(\omega)$ with respect to the oscillating field angular frequency $\omega$. The resonance curves are computed within the ILLG model with $\tau=2\times10^{-10}\ s$, for dimensionless damping $\alpha=0.1$ and for an applied static field $H=2\ T$. Two resonance peaks are observed : the precession resonance at lower angular frequency which is the usual FMR and the nutation resonance at higher angular frequency. Inset : Example of the calculation of $\chi_{\perp}$ such that $<M_{\perp}>=\chi_{\perp} h_{\perp}$ obtained for $\omega=1.2\times 10^{11}\ rad.s^{-1}$. 
}
\label{fig1}
\end{figure}
The inset of Fig.\ \ref{fig1} shows that the response is linear $<M_{\perp}>=\chi_{\perp} h_{\perp}$ wherefrom we extract the transverse susceptibility $\chi_{\perp}$ using a linear fitting. We repeat the same procedure for each oscillating field angular frequency $\omega$ which gives the resonance curve $\chi_{\perp}(\omega)$ of the transverse susceptibility shown in Fig.\ \ref{fig1}. Two peaks clearly appear, the usual FMR peak associated to the precession velocity, and the nutation peak associated to the nutation dynamics originating from the inertial term.

\section{Results}
\label{Results}
\subsection{Effects of $\tau$}

\begin{figure}[!h]
 \centering
\includegraphics[scale=0.54]{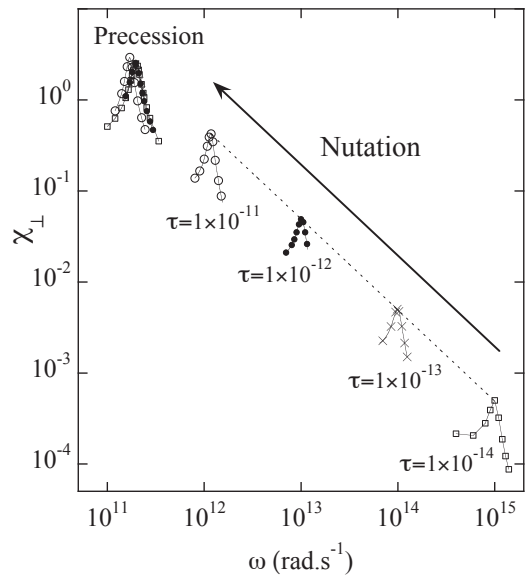}
\caption{Resonance curves of the transverse susceptibility showing the displacement of the nutation peak  caused by the variation of $\tau$ : $\tau=10^{-11}$ s (open circles), $10^{-12}$ s (filled circles), $10^{-13}$ s (crosses), and $10^{-14}$ s (open squares). These curves are simulated using the ILLG model with 
$ \alpha=0.1$, and $H=2T$. Note that the precession peak positions are only slightly affected. The dotted line shows the power law fitted on  
 $\chi_{\perp} \propto 1/\omega_\textrm{nu}^\textrm{\tiny ILLG}$, where $\omega_\textrm{nu}^\textrm{\tiny ILLG}$ is the resonance angular frequency of the nutation. 
}
\label{fig2}
\end{figure}

We now examine the ILLG model when varying the characteristic time $\tau$.
For different values of the parameter $\tau$,  we show in Fig.\ \ref{fig2} the typical profiles of the transverse susceptibility $\chi_{\perp}$ versus the angular frequency $\omega$ of the applied oscillating field. The four resonance curves plotted in figure \ref{fig2} are obtained by numerical simulations with $H=2\ T$ and $\alpha=0.1$.
They show how the nutation resonance peak position depend on the value of $\tau$. As $\tau$ is increased, the nutation peak moves towards 
the precession peak with an increasing intensity which is an order of magnitude smaller than the precession one for $\tau=10^{-11}\ s$. Note that the transverse susceptibility at the resonance follows a power law of the form $\chi_{\perp}(\omega_\textrm{nu}^\textrm{\tiny{ILLG}}) \propto 1/\omega_\textrm{nu}^\textrm{\tiny{ILLG}}$, where $\omega_\textrm{nu}^\textrm{\tiny{ILLG}}$ is defined as the nutation resonance angular frequency. A similar power law is reported for the precession peak obtained for different static fields $H$ (see section \ref{Scaling}). 

We now compare the analytical and numerical simulation results concerning the positions in the angular frequency domain of both the precession and nutation resonance peaks.
\subsubsection{Precession peak}

\begin{figure}[!h]
 \centering
\includegraphics[scale=0.4]{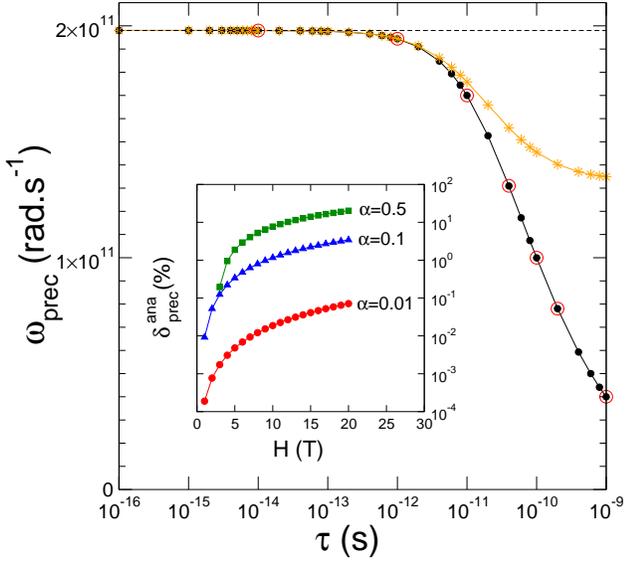}
\caption{(Color online) Comparaison of the analytical and numerical simulation results for the precession angular frequency obtained for $\alpha=0.1$ and $H=2\ T$. Filled circles (black) are the precession angular frequency $\omega_\textrm{prec}^\textrm{exact}$, open circles (red) are the position of the precession resonance peaks $\omega_\textrm{prec}^\textrm{\tiny ILLG}$, stars (orange) are the approximate precession angular frequencies $\omega_\textrm{prec}^\textrm{approx}$ valid for small values of $\tau$. The dashed line (black) is the LLG precession angular frequency, {\it i.e.} without inertial term. Inset : relative difference $\delta_\textrm{prec}^\textrm{ana}$  for three different dampings.
}
\label{fig3}
\end{figure}

We define $\omega_\textrm{prec}=|\dot\phi_\textrm{prec}|$ as the angular frequency of the precession.
When computed from the exact expressions (\ref{eq9}) and (\ref{eq10}) we will refer to $\omega_\textrm{prec}^\textrm{exact}$, and when computed from the approximate expression (\ref{eq11}) we will refer to $\omega_\textrm{prec}^\textrm{approx}$. Finally, we will denote by $\omega_\textrm{prec}^\textrm{\tiny ILLG}$ the angular frequency of the precession resonance peak obtained in the numerical simulations of the ILLG model. 
Eq. (\ref{eq9}) may be easily numerically solved to find the solution $\dot\phi_\textrm{prec}$ for several values of $\alpha$ and $\tau$. 
The behavior with respect to $\tau$ of $\omega_\textrm{prec}$ obtained either analytically or from the simulated FMR curves is shown in Fig.\ \ref{fig3}.
There is an excellent agreement between the analytical prediction $\omega_\textrm{prec}^\textrm{exact}$ and the precession resonance peak $\omega_\textrm{prec}^\textrm{\tiny ILLG}$ obtained in  numerical simulations.
 We also show in Fig.\ \ref{fig3} the precession angular frequency $\omega_\textrm{prec}^\textrm{approx}$. For $\tau < 10^{-11}\ s$ and $\alpha=0.1$, it nicely agrees with the exact analytical value and with the numerical simulation results, but the approximate solution becomes no longer valid for $\tau > 10^{-11}\ s$. To quantify the validity of the approximate solution we compute, for $\tau=10^{-12}\ s$ and for three different dampings, the relative difference 
$$\delta_\textrm{prec}^\textrm{ana} =\frac{\omega_\textrm{prec}^\textrm{approx}-\omega_\textrm{prec}^\textrm{exact}}{\omega_\textrm{prec}^\textrm{exact}}\times 100$$
We show in the inset of Fig.\ \ref{fig3} the evolution of $\delta_\textrm{prec}^\textrm{ana}$ with respect to the applied static field $H$. For $H<20\ T$  the relative difference remains less than $0.1\%$ for small damping $\alpha=0.01$, and remains less than $3\%$ for moderate damping $\alpha=0.1$. For large damping $\alpha=0.5$ the approximate solution remains valid for small fields, but for $12\ T<H<20\ T$ the error becomes larger than $10\%$.
\subsubsection{Nutation peak}
\label{NutationPeakAnalSimul}

\begin{figure}[!h]
 \centering
\includegraphics[scale=0.4]{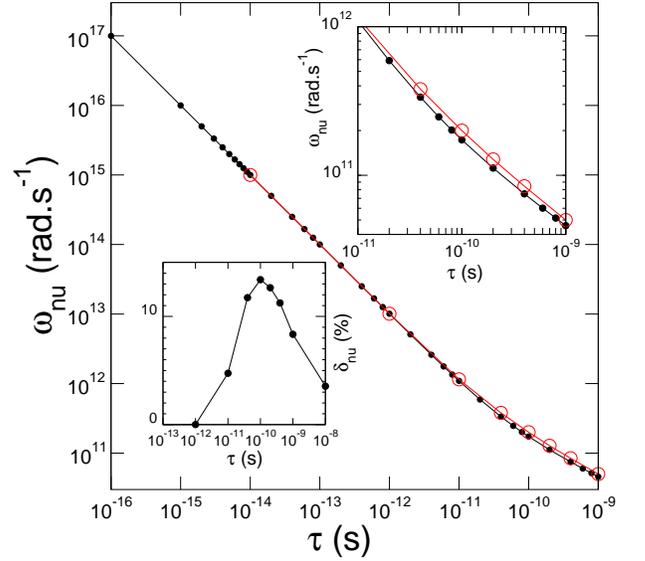}
\caption{(Color online) Comparaison of the analytical and numerical simulation results for the nutation angular frequency obtained for $\alpha=0.1$ and $H=2\ T$. Filled circles (black) are the approximate nutation 
angular frequencies $\omega_\textrm{nu}^\textrm{weak}$ and open circles (red) are the positions $\omega_\textrm{nu}^\textrm{\tiny{ILLG}}$ of the simulated nutation resonance peaks. Upper inset : enlargement showing the effect of the non-linear coupling terms of Eq. (\ref{eq13}). Lower inset : relative difference $\delta_\textrm{nu}$ between  $\omega_\textrm{nu}^\textrm{weak}$ and $\omega_\textrm{nu}^\textrm{\tiny{ILLG}}$.
}
\label{fig4}
\end{figure}

Figure \ref{fig4} displays both the analytical prediction of the nutation 
angular frequency $\omega_\textrm{nu}^{\textrm{weak}}$ given by Eq. (\ref{eq15}) and the angular frequency $\omega_\textrm{nu}^\textrm{\tiny IILG}$ of the nutation resonance obtained in the numerical simulations. 
The agreement is excellent for $\tau<10^{-11}\ s$, and
indicates that the non-linear coupling terms of Eq. (\ref{eq13}) do not significantly shift the angular frequency of the nutation resonance from 
 the approximate angular frequency $\omega_\textrm{nu}^\textrm{weak}$. On the contrary, in the range $10^{-11}\ s<\tau<10^{-8}\ s$, the simulated nutation resonance angular frequency is slightly higher than 
$\omega_\textrm{nu}^\textrm{weak}$, as shown in the upper inset of Fig.\ \ref{fig4}. 
In the lower inset of Fig.\ \ref{fig4} we show the relative difference $\delta_\textrm{nu}$ between the approximate nutation 
angular frequency $\omega_\textrm{nu}^\textrm{weak}$ 
 and the nutation resonance angular frequency $\omega_\textrm{nu}^\textrm{\tiny{ILLG}}$ of the numerical simulations, {\it i. e.} 
$$\delta_\textrm{nu}=\frac{\omega_\textrm{nu}^\textrm{\tiny{ILLG}}-\omega_\textrm{nu}^\textrm{weak}}{\omega_\textrm{nu}^\textrm{\tiny{ILLG}}}\times 100$$ 
We therefore see that in the range $10^{-11}\ s<\tau<10^{-8}\ s$, the 
 approximate nutation angular frequency remains less than $15\%$ close to the simulated nutation resonance angular frequency.\\

\subsection{Scaling and overview of the ILLG equation}
\label{Scaling}

\begin{figure}[!h]
 \centering
\includegraphics[scale=0.4]{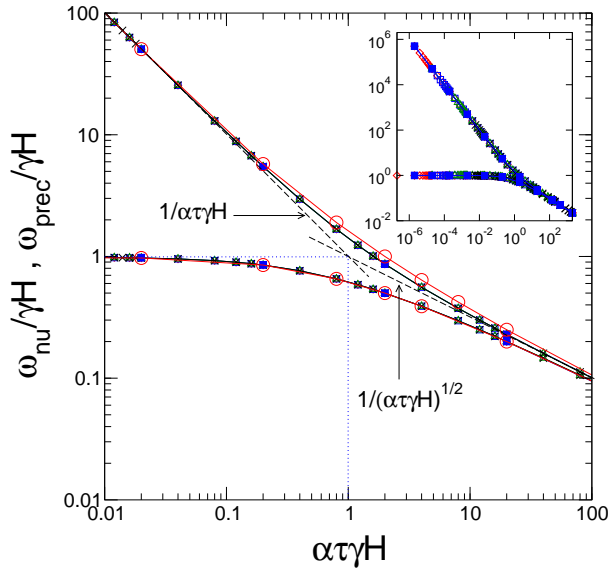}
\caption{(Color online) Scaling curves : nutation  $\omega_\textrm{nu}$ and precession  $\omega_\textrm{prec}$ peak positions in the angular frequency domain divided by $\gamma H$ with respect to $\alpha\tau\gamma H$. Open circles (red) are the nutation and precession resonance peak positions obtained in the numerical simulations for $\alpha=0.1$ and $H=2\ T$. Other points are $\omega_\textrm{nu}^\textrm{weak}$ computed from Eq. (\ref{eq15}), and $\omega_\textrm{prec}^\textrm{exact}$ computed from Eq. (\ref{eq9}). Different values of the static field $H$ and the dimensionless damping $\alpha$ are reported : $H=0.2\ T$ and $\alpha=0.1$ (red open diamonds), $H=2\ T$ (blue open squares for $\alpha=0.1$ and blue filled squares for $\alpha=0.01$), $H=20\ T$ and $\alpha=0.1$ (green open triangles), $H=200\ T$ and $\alpha=0.1$ (black crosses). The dashed lines are the two asymptotic behaviors of the nutation in agreement with Eq. (\ref{eq15}).
Inset : same scaling curves (without red open circles) displayed on larger scales.
}
\label{fig5}
\end{figure}

In the preceding section we investigated the behavior of the ILLG model when varying the characteristic time scale $\tau$ which drives the inertial dynamics. We also vary the static field $H$ and the dimensionless damping $\alpha$. Increasing $H$ moves both the precession and nutation resonance peaks to higher angular frequencies, with smaller and broadened peaks, while increasing the dimensionless damping moves both peaks to lower angular frequencies with still smaller and broadened peaks. Note that the ILLG precession resonances obtained when the static field $H$ is varied show that the transverse susceptibility follows a power law  $\chi_{\perp} \propto 1/\omega_\textrm{prec}^\textrm{\tiny IILG}$ (not shown). This law is the same as the one resulting from the LLG model \cite{Gure}.\\ 
Eq. (\ref{eq15}) suggests a scaling function 
$$\frac{\omega_\textrm {nu}}{\gamma H}=\frac{\sqrt{1+x}}{x}$$
 where $x=\alpha\tau\gamma H$. Scaling curves obtained for different values of $\tau$, $\alpha$ and $H$ are shown in the inset of Fig.\ \ref{fig5} where both the precession and nutation resonance angular frequencies are dispayed with respect to $\alpha\tau\gamma H$. 
Fig.\ \ref{fig5} is an enlargement of the intermediate region of the inset where we added the points obtained by the numerical simulations for $H=2\ T$ and $\alpha=0.1$. The two asymptotic behaviors of the nutation are highlighted with the dashed lines in agreement with Eq. (\ref{eq15}) : when $\alpha\tau\gamma H\ll 1$ then $\omega_\textrm{nu}^\textrm{weak}/\gamma H=1/\alpha\tau\gamma H$, and when $\alpha\tau\gamma H\gg 1$ then $\omega_\textrm{nu}^\textrm{weak}/\gamma H= 1/\sqrt{\alpha\tau\gamma H}$. Remarquably, we see that the precession peak position divided by $\gamma H$ also scales as $\omega_\textrm{prec}/\gamma H\sim 1/\sqrt{\alpha\tau\gamma H}$ when $\alpha\tau\gamma H\gg 1$. The two asymptotic behaviors 
intersect at $\alpha\tau\gamma H=1$ and $\omega/\gamma H=1$. This point corresponds to the maximum value of the LLG precession angular frequency $\omega_\textrm{LLG}/\gamma H=1/(1+\alpha^2)$ which is obtained in the limit case of no damping $\alpha=0$.\\
 The inset of Figure \ref{fig5} indicates that only one resonance peak is expected when $\alpha\tau\gamma H\rightarrow\infty$. In this case, both the nutation and the precession contribute to a unique peak. On the contrary, for finite $\alpha\tau\gamma H$ they remain separated. There are two different well-defined peaks in the investigated range ($\alpha\tau\gamma H\le 100$).
 For $\alpha\tau\gamma H\ll 1$ the precession peak is close to the usual LLG precession peak, and the nutation peak shifts rapidly ($\omega_\textrm{nu}^\textrm{weak}/\gamma H\sim 1/\alpha\tau\gamma H$) to high angular frequencies. In other words, the nutation oscillator defined by Eq. (\ref{eq13}) is independent of the precession for $\alpha\tau\gamma H\ll 1$, whereas both synchronize at the same frequency for $\alpha\tau\gamma H\rightarrow\infty$.\\
 Accurate predictions about the precession and nutation peak positions in the angular frequency domain can be made, as long as the non-linear coupling terms of  Eq. (\ref{eq13}) 
remain weak or compensate each other.

\section{Towards experimental evidence of the inertial dynamics of the magnetization}
\label{Experiments}

\begin{figure}[!h]
 \centering
\includegraphics[scale=0.4]{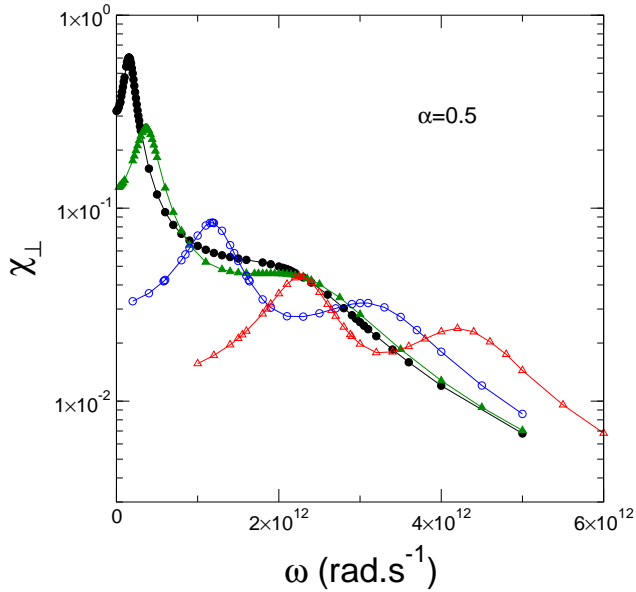}
\caption{(Color online) Resonance curves of the transverse susceptibility $\chi_{\perp}(\omega)$ with respect to the oscillating field angular frequency $\omega$. The resonance curves are computed within the ILLG model with $\tau=10^{-12}\ s$, for a large dimensionless damping $\alpha=0.5$ and for various applied static fields : $H=2\ T$ (black filled circles), $H=5\ T$ (green filled triangles), $H=20\ T$ (blue open circles) and $H=50\ T$ (red open triangles). Both resonance peaks clearly appear for $H=20\ T$ and $H=50\ T$. For $H=5\ T$ and $H=2\ T$ the nutation resonance peak dissapears due to the large damping.
}
\label{fig6}
\end{figure}

\begin{figure}[!h]
 \centering
\includegraphics[scale=0.4]{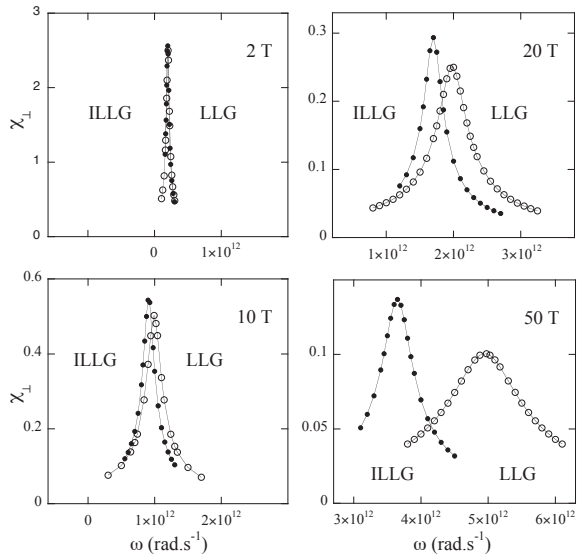}
\caption{Precession resonance curves of the transverse susceptibility simulated for different values of the static field  $H=2\ T$, $10\ T$, $20\ T$, and $50\ T$. ILLG model (filled circles) and non-inertial LLG model (open circles). $\gamma=10^{11}\ rad.s^{-1}.T^{-1}$, $\alpha=0.1$, and $\tau=10^{-12}\ $s.
}
\label{fig7}
\end{figure}

Throughout the preceding sections we studied the new properties of the inertial dynamics of the magnetization within the ILLG model. We specifically considered the FMR framework where a small perpendicular sinusoidal field is applied implying that the small inclination limit holds. We now focus on possible simple experiments in such FMR framework that should highlight the inertial dynamics of the magnetization.\\
The first direct evidence would of course be the measure of the nutation resonance peak at frequencies larger than the precession resonance peak. Since the expected nutation resonance peak is given by Eq. (\ref{eq15}), the evolution with the static field $H$ may be used to discriminate the nutation resonnce peak from possible other higher frequency peaks.\\ 
However the amplitude of the nutation resonance peak is smaller than for the precession peak, and it may be tricky in unfavorable situations to measure such a peak, for example in materials with small characteristic time $\tau$. Furthermore, for large dimensionless damping $\alpha$ both peaks have smaller amplitude and are rounded. It may even appear that the nutation resonance peak of the magnetization in the ILLG model disappears for a large damping, like the resonant peak of the classical driven damped harmonic oscillator. For example Fig.\ \ref{fig6} shows that for materials with a large damping ($\alpha=0.5$) the resonance peaks are smaller and rounded compared to smaller damping ($\alpha=0.1$), and the nutation resonance peak disappears for $H\le 5\ T$.\\
It is therefore necessary to find measurable characteristics of the magnetization inertial dynamics other than the direct measure of the nutation resonance peak. Actually, we show in the following that beyond the nutation resonance peak, the inertial dynamics has measurable effects on the precession resonance peak. Indeed, as shown in Fig.\ \ref{fig7}, the shape of the precession peak and its position in the angular frequency domain are modified by the inertial dynamics. 
And the effects are shown to be more pronounced for large damping materials and for large static magnetic fields $H$. To show these effects we compare the precession resonance angular frequencies $\omega_\textrm{prec}^\textrm{\tiny ILLG}$ and $\omega_\textrm{prec}^\textrm{\tiny LLG}$ obtained in the numerical simulations of both the ILLG and non-inertial LLG models. We use two different dimensionless damping $\alpha=0.1$ and $\alpha=0.5$, and vary the amplitude $H$ of the static magnetic field. For the ILLG model, we choose, as in Ref.\ \onlinecite{inertia}, a rough estimation of the characteristic time scale $\tau=10^{-12}\ s$. 

\subsection{Angular frequency of the precession resonance peak} 
\begin{figure}[!h]
 \centering
\includegraphics[scale=0.4]{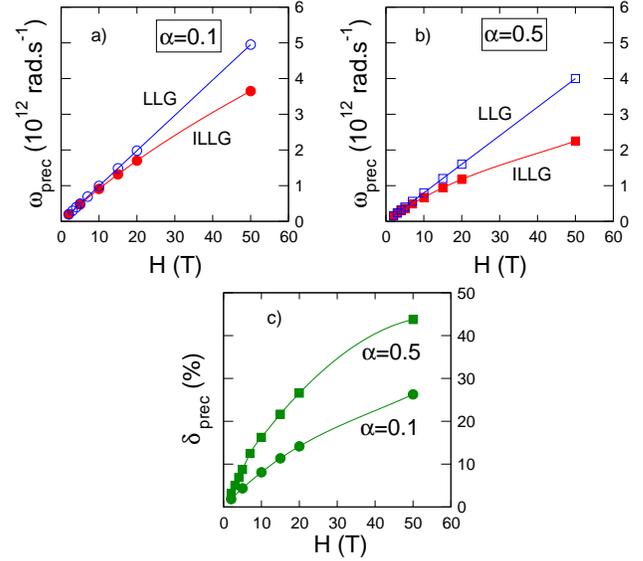}
\caption{(Color online) (a) and (b) Precession resonance angular frequency with respect to the applied static field. Results obtained in the numerical simulations of the ILLG model (with $\tau=10^{-12}\ s$) and non-inertial LLG model, for dimensionless damping (a) $\alpha=0.1$ (blue open circles for LLG and red filled circles for ILLG) and (b) $\alpha=0.5$ (blue open squares for LLG and red filled squares for ILLG). (c) Relative difference $\delta_\textrm{prec}$ between LLG and ILLG precession resonance angular frequencies for $\alpha=0.1$ (green filled circles) and $\alpha=0.5$ (green filled squares).
}
\label{fig8}
\end{figure}

We first look at the position of the precession resonance peak in the angular frequency domain. Fig.\ \ref{fig8}(a) and \ref{fig8}(b) display the evolution of the resonance angular frequency $\omega_\textrm{prec}$ with respect to $H$ obtained for $\alpha=0.1$ and $\alpha=0.5$ within the numerical simulations of both the ILLG and LLG models. 
As expected the resonance angular frequency of the LLG precession is linear with $H$ since $\omega_\textrm{prec}^\textrm{\tiny LLG}=\gamma H/(1+\alpha^2)$ whereas the behavior is not linear in $H$ for the ILLG model. In Fig.\ \ref{fig8}(c) we plot the relative difference $$\delta_\textrm{prec}=\frac{\omega_\textrm{prec}^\textrm{\tiny LLG}-\omega_\textrm{prec}^\textrm{\tiny ILLG}}{\omega_\textrm{prec}^\textrm{\tiny LLG}}\times 100$$ between both resonance angular frequencies. The relative distance between both precession peaks increases with $H$ and with the dimensionless damping $\alpha$. 

\subsection{Width of the precession resonance peak} 
\begin{figure}[!h]
 \centering
\includegraphics[scale=0.4]{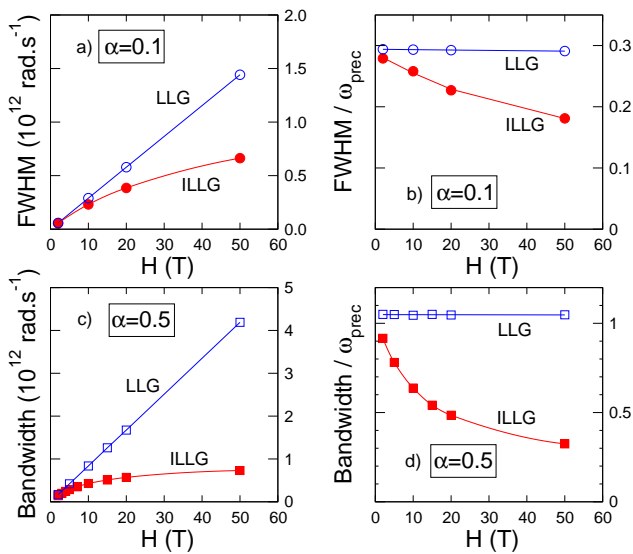}
\caption{(Color online) (a) Full width at half maximum (FWHM) for the precession resonance peak for $\alpha=0.1$ within the LLG (blue open circles) and the ILLG (red filled circles) models. (b) FWHM divided either by $\omega_\textrm{prec}^\textrm{\tiny LLG}$ (blue open circles) or by $\omega_\textrm{prec}^\textrm{\tiny ILLG}$ (red filled circles). (c) Bandwidth of the precession resonance peak for $\alpha=0.5$ within the LLG (blue open squares) and ILLG (red filled squares) models. (d) Bandwidth divided either by $\omega_\textrm{prec}^\textrm{\tiny LLG}$ (blue open squares) or by $\omega_\textrm{prec}^\textrm{\tiny ILLG}$ (red filled squares).\\
The numerical simulations of the ILLG model are computed with $\tau=10^{-12}\ s$
}
\label{fig9}
\end{figure}

We now examine the evolution with $H$ of the shape of the precession resonance peak obtained in the simulations of the ILLG and LLG models. For $\alpha=0.1$, the full width at half maximum (FWHM) is shown in Fig.\ \ref{fig9}(a) while Fig.\ \ref{fig9}(b) displays the FWHM divided by the resonance angular frequency. 
For large damping $\alpha=0.5$ we change the criterion since the reduced amplitude of the resonant peak does not allow anymore to compute the FWHM. We therefore compute the bandwith defined by the width of the peak at $A_\textrm{max}/\sqrt{2}$ where $A_\textrm{max}$ is the maximum value of the peak. The bandwidth for $\alpha=0.5$ is shown in  Fig.\ \ref{fig9}(c) and the bandwidth divided by the resonance angular frequency is plotted in  Fig.\ \ref{fig9}(d).
The numerical simulations of the ILLG and LLG models lead to different behaviors for the shape of the precession resonance peak. In the LLG model the FWHM and the bandwidth exhibit a linear evolution with the applied static field which results in a constant evolution when divided by the resonance angular frequency. Very different behaviors are observed within the ILLG model where no linear evolution of the FWHM or the bandwidth is measured. \\

Figs.\ \ref{fig8} and \ref{fig9} show that high applied static fields in large damping materials produce large differences between the positions and shapes of the precession resonance peaks originating from the LLG and ILLG models. Therefore, applying high static fields in large damping materials better allows to differentiate the precession peak originating from the ILLG and LLG models.\\
Although the theory is clear and allows in principle to differentiate inertial from non-inertial dynamics when examining both precession resonance peaks, the experimental investigations are rather more complex. Indeed, the experimental demonstration of inertial effects first necessitate to identify and control the different contributions to the effective field (anisotropy, dipolar interaction, magnetostriction, ...) other than the applied static field.
\section{Conclusion}
\label{Conclusion}

The magnetization dynamics in the ILLG model that takes into account inertial effects has been studied from both analytical and numerical points of view. Within the FMR context, a nutation resonance peak is expected in addition to the usual precession resonance peak.\\
Analytical solutions of the inertial precession and nutation 
 angular frequencies are presented. The analytical solutions nicely agree with the numerical simulations of the resonance curves in a broad range of parameters. At first, we investigated the effects of the time scale $\tau$ which drives the additional inertial term introduced in Eq. (\ref{eq2}) compared to the usual LLG equation Eq. (\ref{eq1}). We also varied the dimensionless damping $\alpha$ and the static magnetic field $H$, and a scaling function with respect to $\alpha\tau\gamma H$ is found for the nutation  
angular frequency. Remarquably, the same scaling holds for the precession angular frequency when $\alpha\tau\gamma H\gg 1$.\\
In the second part of the paper we focussed on the signatures of the inertial dynamics which could be detected experimentally within the FMR context. We showed that beyond the measure of the nutation resonance peak which would be a direct signature of the inertial dynamics, the precession is modified by inertia and the ILLG precession resonance peak is different from the usual LLG precession peak. Indeed, whereas a linear evolution with respect to $H$ is expected for the LLG precession resonance angular frequency, the ILLG precession resonance angular frequency is clearly non-linear. Furthermore, the shape of the precession resonance peak is different in the LLG and ILLG models. Again, the width variation of the precession resonance peak is non-linear in the ILLG dynamics as opposed to the linear evolution with $H$ in the LLG dynamics. We also showed that the difference between both LLG and ILLG precession peaks is more pronounced when the damping is increased and when $\tau$ is increased. For example the discrepancy between the LLG and ILLG precession resonance angular frequencies at $H = 20\ T$ for $\tau = 1\ ps$ is expected to be of the order of $20\%$ for $\alpha=0.1$ and $30\%$ for $\alpha=0.5$. Therefore, large damping materials are better candidates to experimentally evidence the inertial dynamics of the magnetization.\\
Finally, a specific behavior of the amplitude of the magnetic susceptibility as a function of the nutation resonance angular frequency $\omega_\textrm{nu}$ is predicted, of the form $\chi_{\perp}(\omega_\textrm{nu}) \propto \omega_\textrm{nu}^{-1}$ (analogous to that of the usual FMR susceptibility). This law could be a useful criterion in order to discriminate the nutation peak among the other excitations that could also occur close to the infrared region (100 GHz up to 100 THz) in a ferromagnetic material.


\end{document}